\documentclass[onecolumn,prl, preprint, superscriptaddress]{revtex4-2}
\usepackage{bm}
\usepackage[colorlinks=true,linkcolor=blue,citecolor=blue]{hyperref}
\usepackage{times}
\usepackage{amsmath}
\usepackage{amssymb}
\usepackage{amsthm}
\usepackage{amsfonts}
\usepackage{enumerate}
\usepackage{latexsym}
\usepackage{ifpdf}
\newcommand{\beq}{\begin{equation}}
\newcommand{\eeq}{\end{equation}}
\usepackage{graphicx}
\usepackage{makeidx}
\usepackage{color}

\usepackage[version=4]{mhchem}
\usepackage{textcomp,mathcomp}
\begin{document}

\title{Mean field magnetism and spin frustration in a double perovskite oxide with compositional complexity}

\author {Nandana Bhattacharya}
\email{nandanab@iisc.ac.in}
\affiliation  {Department of Physics, Indian Institute of Science, Bengaluru  560012, India}
\author {Ravi Kiran Dokala}
\affiliation  {Department of Materials Science and Engineering, Uppsala University, Box 35, SE-751 03,
Uppsala, Sweden}
\author{Sourav Chowdhury}
\affiliation {Deutsches Elektronen-Synchrotron DESY, 22607 Hamburg, Germany}
\author {Suresh Chandra Joshi}
\affiliation  {Department of Physics, Indian Institute of Science, Bengaluru  560012, India}
\author {Subha Dey}
\affiliation  {Department of Physics, Indian Institute of Science, Bengaluru  560012, India}
\author{Jayjit Kumar Dey}
\affiliation {Deutsches Elektronen-Synchrotron DESY, 22607 Hamburg, Germany}
\author{Subhajit Nandy}
\affiliation {Deutsches Elektronen-Synchrotron DESY, 22607 Hamburg, Germany}
\author{Daniel P\'erez Salinas}
\affiliation{ALBA Synchrotron Light Source, Cerdanyola del Valles, Barcelona E-08290, Spain}
\author{Manuel Valvidares}
\affiliation{ALBA Synchrotron Light Source, Cerdanyola del Valles, Barcelona E-08290, Spain}
\author{Moritz Hoesch}
\affiliation {Deutsches Elektronen-Synchrotron DESY, 22607 Hamburg, Germany}
\author{Roland Mathieu}
\affiliation {Department of Materials Science and Engineering, Uppsala University, Box 35, SE-751 03,
Uppsala, Sweden}
\author {Srimanta Middey}
\email{smiddey@iisc.ac.in}
\affiliation  {Department of Physics, Indian Institute of Science, Bengaluru 560012, India}

\begin{abstract}

 The rise of high-entropy oxides as a major functional materials design principle in recent years has prompted us to investigate how compositional disorder affects long-range magnetic ordering in double perovskite oxides. Since ferromagnetic insulators are emerging as an important platform for lossless spintronics, we consider the $RE_2$NiMnO$_6$ ($RE$ : rare-earth) family and investigate single-crystalline films of (La$_{0.4}$Nd$_{0.4}$Sm$_{0.4}$Gd$_{0.4}$Y$_{0.4}$)NiMnO$_{6}$ grown on SrTiO$_3$ (001) substrates  in this work. Despite configurational disorder and high cationic size variance at the $RE$ site, the material exhibits robust ferromagnetic ordering with a Curie temperature ($T_\mathrm{c}$) of approximately 150 K. This $T_\mathrm{c}$ is consistent with the expectation based on consideration of the average ionic radii of the rare-earth ($RE$) sites in the bulk $RE_2$NiMnO$_6$.
 Below $T_\mathrm{c}$, Raman spectroscopy measurement finds a deviation from anharmonic behavior, where the phonon renormalization aligns with a mean-field approximation of spin-spin correlation. 
 At lower temperature magnetic $RE$ ions also contributed to the magnetic behavior and the system
  displays a reentrant spin-glass-like behavior. This study demonstrates that while a mean-field approach serves as a viable starting point for predicting the long-range transition temperature, microscopic details of the complex magnetic interactions are essential for understanding the low-temperature phase.

\end{abstract}

\maketitle

\begin{figure*} 
	\vspace{-2pt}
	\hspace{-2pt}
	\includegraphics[width=1.0\textwidth]{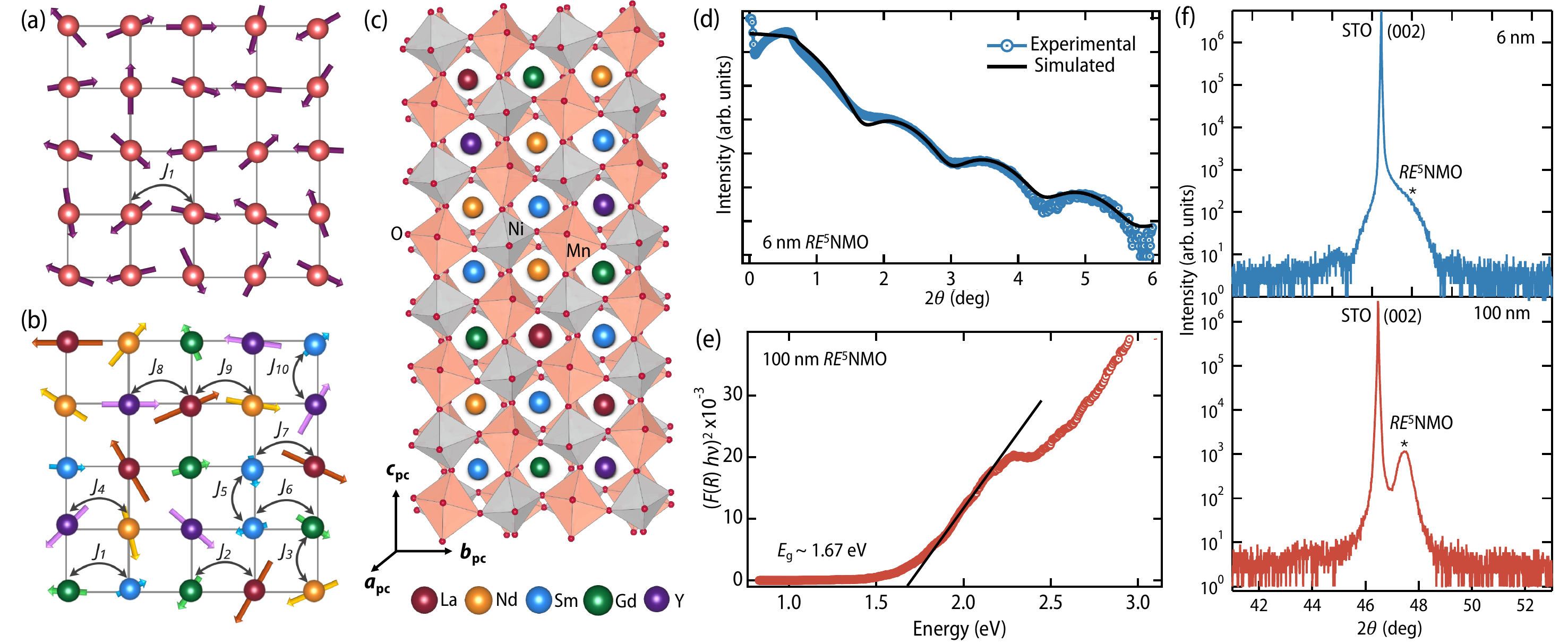}
	\caption{\label{Fig1} {{\bf  Film characterization and electronic structure:} (a) Schematic depicting the exchange coupling in a square lattice with (a) one kind of atom exhibiting uniform nearest neighbour interactions (b) random distribution of five different kinds of atoms with different nearest neighbour couplings. (c) Crystal structure of an Ni-Mn ordered $RE^5$NMO where the $A$ site is shown for a random distribution of 5 $RE$ cations [Legend for colors representing each $RE$ cation has been shown at the bottom].(d) XRR experimental data and fitting for 6 nm $RE^5$NMO film on STO. (e) The reflectance data were used to derive the Kubelka-Munk function, which is an absorption equivalent commonly used for the diffused reflectance mode given by,
$F$($R$) = (1-$R^2$) / 2$R$, where $R$ is the reflectance ~\cite{Lopez:2012p1}. The Tauc relation using the reflectance mode is given by $[\left[F(R) \cdot h\nu\right]^n = A(h\nu - E_g)$]
where, $F$($R$) is the Kubelka-Munk function described above, $A$ is a characteristic constant independent of the photon energy, $h\nu$ is the photon energy and $E_g$ is the band gap ~\cite{Tauc:1966p627}. We use $n$ = 2 here as $RE_2$NiMnO$_6$ exhibit direct optical band gaps ~\cite{Arima:1993p17006, Yi:2022p979}. The black line is the linear extrapolation for the Tauc plot to find the band gap.
    (f) XRD for 6 nm (upper panel) and 100 nm (lower panel) $RE^5$NMO film on STO, * denotes the film peak. 
    }}
\end{figure*}

Transition metal oxides are a fertile ground for a wide array of magnetic phases, from long-range ordered phases like ferromagnetism, antiferromagnetism ~\cite{Coey:2010}, altermagnetism ~\cite{Feng:2022p735,Song:2025p1} to frustrated states such as spin-glass~\cite{Binder:1986p801} and spin-liquid ~\cite{Broholm:2020p0668}. This rich magnetism, along with its coupling to underlying lattice, orbital, and charge degrees of freedom, give rise to fascinating magnetic phenomena such as colossal magnetoresistance~\cite{Tokura:2006p797}, skyrmions~\cite{Wang:2018p1087}, anomalous Hall effect~\cite{Nagaosa:2010p1539}, topological Hall effect, etc~\cite{Vistoli:2019p67, Ojha:2020p2000021}.  Modern condensed matter physics often focuses on engineering these properties in thin film form by controlling strain, confinement, heterointerfaces, geometrical lattice engineering, etc~\cite{Bhattacharya:2014p65, Hellman:2017p025006}. In recent years, high entropy oxides (HEOs) containing five or more number of elements are being investigated as a new paradigm of materials design, providing several advantages over traditional materials~\cite{Rost:2015p8485, Mazza:2024p230501,Aamlid:2023p5991,Han:2024p846, Schweidler:2024p266}. While conventional notion suggests that such compositional disorder would disrupt magnetic interactions and favor a spin-glass-like phase or inhomogeneous magnetic phase~\cite{Witte:2019p034406, Mazza:2022p2200391,Mazza:2021p094204,Sarkar:2023p2207436,Ke:2024p2312856}, a surprising number of magnetic HEOs exhibit long-range magnetism~\cite{Zhang:2019p3705, Jimenez:2019p122401, Nevgi:2025p7038,Min:2024p24320, Clulow:2024p6616,Pramanik:2024p3382}. This raises a fundamental question: How can we describe the magnetism in these systems? Instead of focusing on microscopic variations [Fig.~\ref{Fig1}(a), (b)], could a mean-field approach~\cite{Blundell:2001}, which considers an average internal magnetic field around each magnetic moments, be a suitable starting point for understanding their magnetic behavior?  To examine this, we focus on an insulating double perovskite oxide in this work.

Double perovskite oxides (DPOs) with the general formula $A_2$$BB$$'$O$_6$ (where $B, B$' are transition metal (TM) cations) show a variety of magnetic ground states owing to their structural flexibility and the possibility of incorporating two distinct TM cations~\cite{Vasala:2015p1}. 

In this work, we consider the $RE_2$NiMnO$_6$ family ($RE$=La, Pr, Nd,...Y), which shows a transition from paramagnetic insulating to ferromagnetic insulating phase upon lowering the temperature~\cite{Rogado:2005p2225,Choudhury:2012p127201,Pal:2019p045122,Su:2015p13260, Nasir:2019p141}.  Due to their potential for use in low-power, energy-efficient spintronic devices, ferromagnetic insulators (FMIs) are a subject of significant research interest. This has led to a push for the development of transition-metal oxide–based FMIs, facilitated by recent advances in oxide thin-film growth techniques~\cite{Li:2023p3638,Li:2025p016702}.
 In the $RE_2$NiMnO$_6$  series of FMIs, the ferromagnetic ordering is caused by Ni$^{2+}$-O-Mn$^{4+}$ ferromagnetic superexchange, and the transition temperature ($T_\mathrm{c}$) strongly depending on the choice of $RE$-ion (Table~\ref{t1})~\cite{Booth:2009p1559}. The magnetic interactions between the $RE$-sites and the Ni/Mn sublattice have also been reported~\cite{Sanchez:2011226001, Pal:2019p045122, Booth:2009p1559}. To examine the effect of compositional complexity under a high-entropy setting, we examine
(La$_{0.4}$Nd$_{0.4}$Sm$_{0.4}$Gd$_{0.4}$Y$_{0.4}$)NiMnO$_6$ $(RE^5$NMO) [See Fig.~\ref{Fig1}(c) for schematic depiction]. The choice of this particular combination of $RE$ ions is motivated to obtain a high cationic variance at the $RE$ site ($\sigma^2= \sum_{i=1}^{n}y_i r_i^2 - <r_A>^2$  $\sim$ 23.3 pm$^2$, where $r_i$ is the cationic radius with fractional occupancy $y_i$, and $<r_A>$ is the mean radius~\cite{Mazza:2023p013008}).

\begin{table*}
	\centering
		\caption{Electronic configuration, magnetic moments and ionic radii of $RE^{3+}$ cations and corresponding magnetic transitions in $RE_2$NiMnO$_6$ compounds. $g_J$ values adapted from Ref.~\cite{Booth:2009p1559}.}
		\begin{tabular}{c  c  c  c  c  c  c  c  c }
		\hline
		$RE^{3+}$ & Electronic  & $L$ & $S$ & Ground-state  & Effective $RE$ moment  & Ionic  & $RE_2$NiMnO$_6$ & $T_\mathrm{c}$ (K)~\cite{Booth:2009p1559}\\
		& config.& &  & term $^{2S+1}L_{J}$ & $\mu_J = \sqrt{J(J+1)}\,\mu_B$ &radii ({\AA}) ~\cite{Shannon:1976p751}& \\
\hline\hline
		La$^{3+}$ & 4$f^0$ & 0 & 0 & $^{1}S_{0}$ & 0 & 1.160  & La$_2$NiMnO$_6$ & 270 \\
		Pr$^{3+}$ & 4$f^2$ & 5 & 1 & $^{3}H_{4}$ & 3.58 & 1.126 & Pr$_2$NiMnO$_6$ & 212 \\
		Nd$^{3+}$ & 4$f^3$ & 6 & 3/2 & $^{4}I_{9/2}$ & 3.64 & 1.109 & Nd$_2$NiMnO$_6$ & 194 \\
		Sm$^{3+}$ & 4$f^5$ & 5 & 5/2 & $^{6}H_{5/2}$ & 0.84 & 1.079 & Sm$_2$NiMnO$_6$ & 157 \\
		Eu$^{3+}$ & 4$f^6$ & 3 & 3 & $^{7}F_{0}$ & 0 & 1.066 & Eu$_2$NiMnO$_6$ & 143 \\
		Gd$^{3+}$ & 4$f^7$ & 0 & 7/2 & $^{8}S_{7/2}$ & 7.94 & 1.053 & Gd$_2$NiMnO$_6$ & 128 \\
		Tb$^{3+}$ & 4$f^8$ & 3 & 3 & $^{7}F_{6}$ & 9.72 & 1.040 & Tb$_2$NiMnO$_6$ & 111 \\
		Dy$^{3+}$ & 4$f^9$ & 5 & 5/2 & $^{6}H_{15/2}$ & 10.63 & 1.027 & Dy$_2$NiMnO$_6$ & 93 \\
		Ho$^{3+}$ & 4$f^{10}$ & 6 & 2 & $^{5}I_{8}$ & 10.60 & 1.019 & Ho$_2$NiMnO$_6$ & 82 \\
		Y$^{3+}$ & 4$d^0$ & 0 & 0 & $^{1}S_{0}$ & 0 & 1.015 & Y$_2$NiMnO$_6$ & 79 \\
		\hline
		\end{tabular}
	\label{t1}
\end{table*}

In this study, single crystalline films of $RE^5$NMO were grown on SrTiO$_3$ (001) substrates via pulsed laser deposition (PLD). DC magnetometry measurements on a 100 nm film revealed a robust ferromagnetic transition with a $T_\mathrm{c}$ of approximately 150 K. This $T_\mathrm{c}$ is notably close to that reported for Sm$_2$NiMnO$_6$~\cite{Booth:2009p1559}, implying that the net ferromagnetic exchange is primarily governed by the average bond angles which, in turn, are controlled by the size of the $RE$ ions. Raman spectroscopy revealed a clear deviation in the phonon frequency from the anharmonic model observed below $T_\mathrm{c}$. Intriguingly, this deviation, occurring just below $T_c$, could be accounted considering a mean-field model of magnetism~\cite{Iliev:2007p151914}. However, a distinct magnetic anomaly emerged near 35 K, below which the mean-field model failed to describe the system's behavior and  the system exhibited reentrant spin glass-like behavior. 
 Although both Ni-Mn antisite disorder (ASD) as well as $RE$ site magnetic disorder are deemed to be the drivers of this spin glass state, the transition temperature is markedly lower than the ASD driven transitions reported in parent $RE$ compounds~\cite{Shi:2011p245405,Iliev:2009p023515,Choudhury:2012p127201,Singh:2016p74} and surprisingly matches the $RE$ interaction temperature scale.

\section*{Results}

{\color{magenta}\bf Film growth and characterizations:}
${RE}^5$NMO films have been grown on SrTiO$_3$ [001] (STO) substrates by a Neocera PLD system (growth parameter details are in Methods section). The structural quality of the films have been verified by X-ray reflectivity (XRR) and X-ray diffraction (XRD) measurements using a laboratory based diffractometer. The presence of prominent intensity oscillations across the entire scan range in XRR pattern [Fig.~\ref{Fig1}(d)] is indicative of a highly crystalline nature with a smooth interface. The fitting results yielded a film thickness $\sim$ 6 nm and a film/substrate roughness of about 3.7 {\AA}. The laser pulse calibration obtained from the above analysis was further used for the growth of the 100 nm film.  Fig.~\ref{Fig1}(f) shows the XRD patterns for the 6 nm and 100 nm films, both of which contain a broad film peak next to the sharp substrate peak, confirming their single-crystallinity. While the overlap of the substrate and film prohibits the estimation of the out-of-plane pseudocubic lattice parameter ($c_\mathrm{pc}$) for the 6 nm film, it is found to be $\sim$ 3.83 {\AA} for the 100 nm film. We further characterized the film through the determination of the optical band gap by measuring the diffuse reflectance spectrum and the Tauc plot analysis~\cite{Tauc:1966p627}. The optical band gap $E_{g}$ is found to be $\sim$ 1.67 eV [Fig.~\ref{Fig1}(e)] for the 100 nm film,  which is comparable to the parent members of the $RE_2$NiMnO$_6$ family~\cite{Nasir:2019p141}.

\begin{figure*} 
	\vspace{-2pt}
	\hspace{-2pt}
	\includegraphics[width=0.98\textwidth]{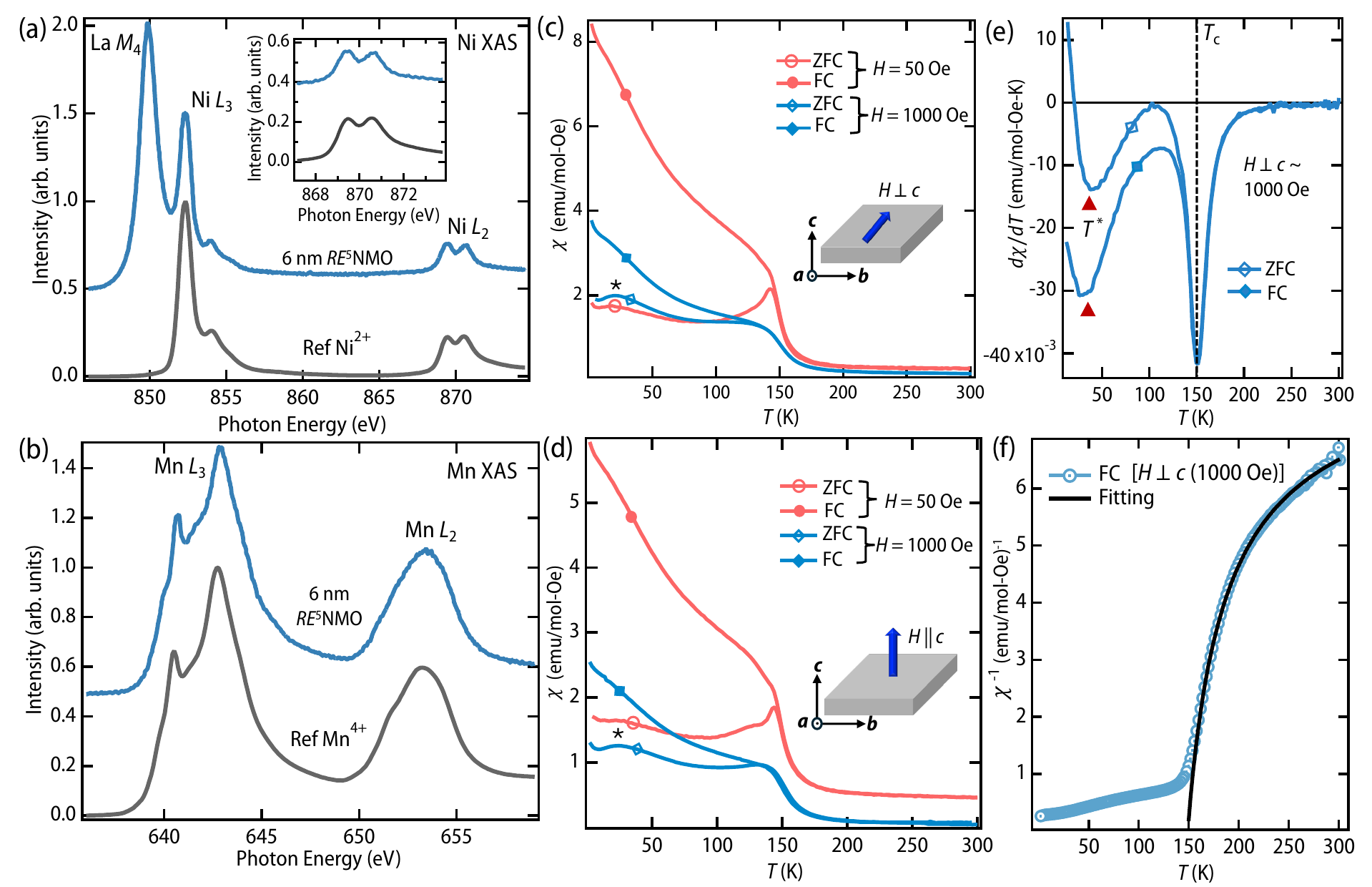}
	\caption{\label{Fig2} {{\bf Long range magnetic ordering:}
    XAS spectra for (a) Ni $L_{3,2}$ edge; Inset : zoomed $L_2$ edge highlighting the characteristic Ni$^{2+}$ splitting, and (b) Mn $L_{3,2}$ edge. The reference for Ni$^{2+}$ and Mn$^{4+}$ have been shown for ease of comparison adapted from Ref. ~\cite{Bhattacharya:2025p176201} corresponding to the spectra for a 20 uc Nd$_2$NiMnO$_6$ film on NdGaO$_3$ substrate. $\chi$ vs $T$ for both ZFC and FC conditions recorded in the warming and cooling cycles respectively, under fields of 50 Oe and 1000 Oe applied (c) in-plane to the film ($H \perp c$) (d) out-of-plane to the film ($H$ $\parallel$ $c$), where $c$ is the out-of-plane crystallographic direction. The schematic of the measurement geometry have been shown in the inset. (e) Temperature derivative of susceptibility, $d\chi/dT$, under 1000 Oe ($H \perp \mathrm{c}$), highlighting the Curie temperature ($T_\mathrm{c}$) and lower temperature anomaly (marked by arrow, $T^*$). (f) $\chi^{-1}$ as a function of temperature for $H \perp c$ under 1000 Oe, fitted with a modified Curie–Weiss formula (black curve).
    }}
\end{figure*}

The origin of ferromagnetism in the parent $RE_2$NiMnO$_6$ series has been explained by the dominance of the ferromagnetic (FM) superexchange interaction between  Ni$^{2+}$-$e_g$ and Mn$^{4+}$-$e_g$ states over the antiferromagnetic (AFM) superexchange between half-filled Ni$^{2+}$-$e_g$ and Mn$^{4+}$-$t_{2g}$ orbitals~\cite{Das:2008p186402}. 
The presence of unwanted Ni$^{3+}$ and Mn$^{3+}$ can give rise to additional magnetic transition at lower temperature~\cite{Shi:2011p245405}. 
 To examine this, we performed the X-ray absorption spectroscopy experiments on Ni $L_{3,2}$ [Fig.~\ref{Fig2}(a)] and Mn $L_{3,2}$ [Fig.~\ref{Fig2}(b)] edges on our 6 nm film at P04 beamline, PETRA III, DESY, using total electron yield (TEY) mode. Reference spectra of a 20 uc Nd$_2$Ni$^{2+}$Mn$^{4+}$O$_6$ film grown on NdGaO$_3$  have also been shown for comparison~\cite{Bhattacharya:2025p176201}.  The Ni $L_3$ spectra are also overlapped with the La $M_4$ spectral feature. As shown in Fig.~\ref{Fig2}(a) and (b), the Ni and Mn spectral features of the film clearly confirm +2 and +4 oxidation states respectively,  similar to oxidation states reported for the bulk parent compounds of $RE_2$NiMnO$_6$ ($RE$ = La, Nd, Sm, Gd and Y) ~\cite{Nasir:2019p141,Pal:2019p045122,Pal:2018p165137}.
Recent studies of Nd$_2$NiMnO$_6$ films grown on SrTiO$_3$ (001) substrates have reported a change in the oxidation state of Mn due to polar catastrophe, while Ni remains in its +2 oxidation state ~\cite{Bhattacharya:2025p176201,Spring:2023p104407}. Such effects are insignificant in the film thickness regime we are probing in this work.

{\color{magenta}\bf Ferromagnetic ordering and validity of a mean field model:}
After confirming the desired oxidation state of Ni and Mn, we focus on investigating the magnetic properties. The magnetic superexchange coupling ($J$) is dependent on the Ni$^{2+}$-O-Mn$^{4+}$ bond angle ($J$ $\propto$ $\cos^2\theta$)~\cite{Das:2008p186402}, which in turn is dependent on the ionic radii of the $RE$ cations leading to a correlation between the $RE$ ionic radii and magnetic transition temperature $T_\mathrm{c}$~\cite{Booth:2009p1559}. To probe the effect of compositional complexity at the $RE$ site on the magnetism, we employed SQUID magnetometry on the 100 nm film. The measurements have been performed in both in-plane (perpendicular to the crystallographic $c$-axis, \emph{H} $\perp$ $c$) and out-of-plane directions (parallel to $c$ - axis, \emph{H} $\parallel$ $c$) with respect to the film geometry [Inset Fig.~\ref{Fig2}(c), (d) respectively]. Fig.~\ref{Fig2} (c) and (d) show the zero field cooled (ZFC) and field cooled (FC) dc susceptibility ($\chi$ = $M$/$H$) measured in warming cycle and cooling cycle, respectively, for $RE^5$NMO under magnetic fields of 50 Oe and 1000 Oe. A clear bifurcation between the FC and ZFC magnetization curves is observed around 150 K, particularly prominent for the lower field (50 Oe). Such behavior is indicative of the onset of long-range ferromagnetic ordering~\cite{Mugiraneza:2022p95,Pal:2019p045122}. The derivative of susceptibility ($d\chi/dT$) also find the $T_\mathrm{c} \sim$ 150 K [Fig.~\ref{Fig2}(e)].

 Amongst the parent counterparts, in Nd$_2$NiMnO$_6$, Sm$_2$NiMnO$_6$, and Gd$_2$NiMnO$_6$, interactions between the $RE$ moments and the Ni/Mn sublattice give rise to distinct low temperature anomalies in the FC $M$–$T$ curves ~\cite{Pal:2019p045122, Booth:2009p1559}. Interestingly, the ZFC magnetization curves exhibit a broad, cusp-like anomaly marked as $*$ in Fig.~\ref{Fig2}(c), (d) for both measurement orientations [also observed in $d\chi/dT$ $\sim$ 35 K marked as $T^*$ in Fig.~\ref{Fig2}(e)]. Such a feature is often associated with the onset of a glassy state, where competing magnetic interactions lead to magnetic frustration~\cite{Guo:2006p262503, Mugiraneza:2022p95, Choudhury:2012p127201}, which shall be demonstrated and discussed in greater details in latter section of this paper.

 In Fig.~\ref{Fig2}(f), we show the fitting
of the inverse of susceptibility using a modified Curie-Weiss model assuming a non-interacting nature of the $RE$ moments at higher temperature: \[\chi(T) = \chi_0 + \frac{C_{{RE}}}{T} + \frac{C_{\text{Ni-Mn}}}{T - \theta_{\text{CW}}}
\]
where $\chi_0$ is the Van-Vleck paramagnetic susceptibility, $C_{RE}$ and $C_\mathrm{Ni-Mn}$ are the Curie constants for the $RE$ and the Ni/Mn sublattices respectively, and $\theta_\mathrm{CW}$ is the Curie-Weiss temperature~\cite{Booth:2009p1559, Blundell:2001}. We fixed the value of \( C_{{RE}} \) during the fitting procedure by estimating the average effective magnetic moment of the $RE$ sublattice using the root-mean-square formula:
$\mu_{\text{eff,$\mathrm{RE}$}} = \sqrt{ \sum_i x_i \mu_{i}^2 }$ ($x_i$ is the atomic fraction and $\mu_{i}$ is the effective moment of the i$^{th}$ $RE$ atom)  which yields \( \mu_{\text{eff,${RE}$}} \approx 3.9\ \mu_B \).
This value was then related to the Curie constant via the relation \(\mu_{\text{eff}} = \sqrt{8C_{{RE}}}\)~\cite{Mugiraneza:2022p95}. From the relation $\mu_{\text{eff}} = \sqrt{\sum g_i^2 S_i (S_i + 1)}$, for Ni$^{2+}$ ($d^8$, $S$= 1) and Mn$^{4+}$ ($d^3$, $S$ =3/2), assuming $g$ = 2 and $g$ = 2.5 for Mn and Ni, respectively~\cite{Booth:2009p1559,Carlin:2012}, the effective moment for the Ni-Mn network should be 5.24 $\mu_B$. Our analysis yields $C_\mathrm{Ni-Mn}$ $\sim$ 5.98$\mu_B$, close to the expected value. Furthermore, a $\theta_\mathrm{CW}$ $\sim$ 149 K is obtained, very close to the $T_\mathrm{c}$ found from the results discussed in Fig.~\ref{Fig2}(c), (d).
The positive value of $\theta_\mathrm{CW}$ and the fact that $T_\mathrm{c}$ $\approx$ $\theta_\mathrm{CW}$ further corroborate that the magnetic transition in our system is ferromagnetic in nature, without any significant role of magnetic frustration effect at higher temperature. Most importantly, the $T_\mathrm{c}$ and $\theta_\mathrm{CW}$ are closer to that of parent Sm$_2$NiMnO$_6$, which share the similar average tolerance factor. These results affirm that the ferromagnetic transition temperature, despite the compositional complex setting at the $RE$ sublattice with large variance, is governed in a mean-field way. This is further examined through Raman spectroscopy.

{\color{magenta}\bf Correlating magnetism to phonon vibrations:} In FMI DPOs, the ferromagnetic ordering has shown to exhibit a pronounced effect on the phonon-frequencies below $T_\mathrm{c}$ owing to strong spin-phonon coupling ~\cite{Iliev:2007p104118,Iliev:2007p151914,Nasir:2019p141}. In the backdrop of possibility of having different types of $RE$ ions surrounding a particular Ni/Mn site, we next investigate the structure-magnetism correlation using Raman spectroscopy as a function of temperature (measurement details are in Method section).  For the entire range of temperature (300 K to 4.2 K), we observe two prominent Raman modes in the range of 400 to 800 cm$^{-1}$ phonon frequencies: $\sim$ 510 and 653 cm$^{-1}$,  consistent with previous reports for monoclinic $RE_2$NiMnO$_6$ ~\cite{Macedo:2015p075201,Iliev:2007p151914, Nasir:2019p141}, also affirming the absence of any structural transitions. 

\begin{figure} 
	\vspace{-2pt}
	\hspace{-2pt}
	\includegraphics[width=0.8\textwidth] {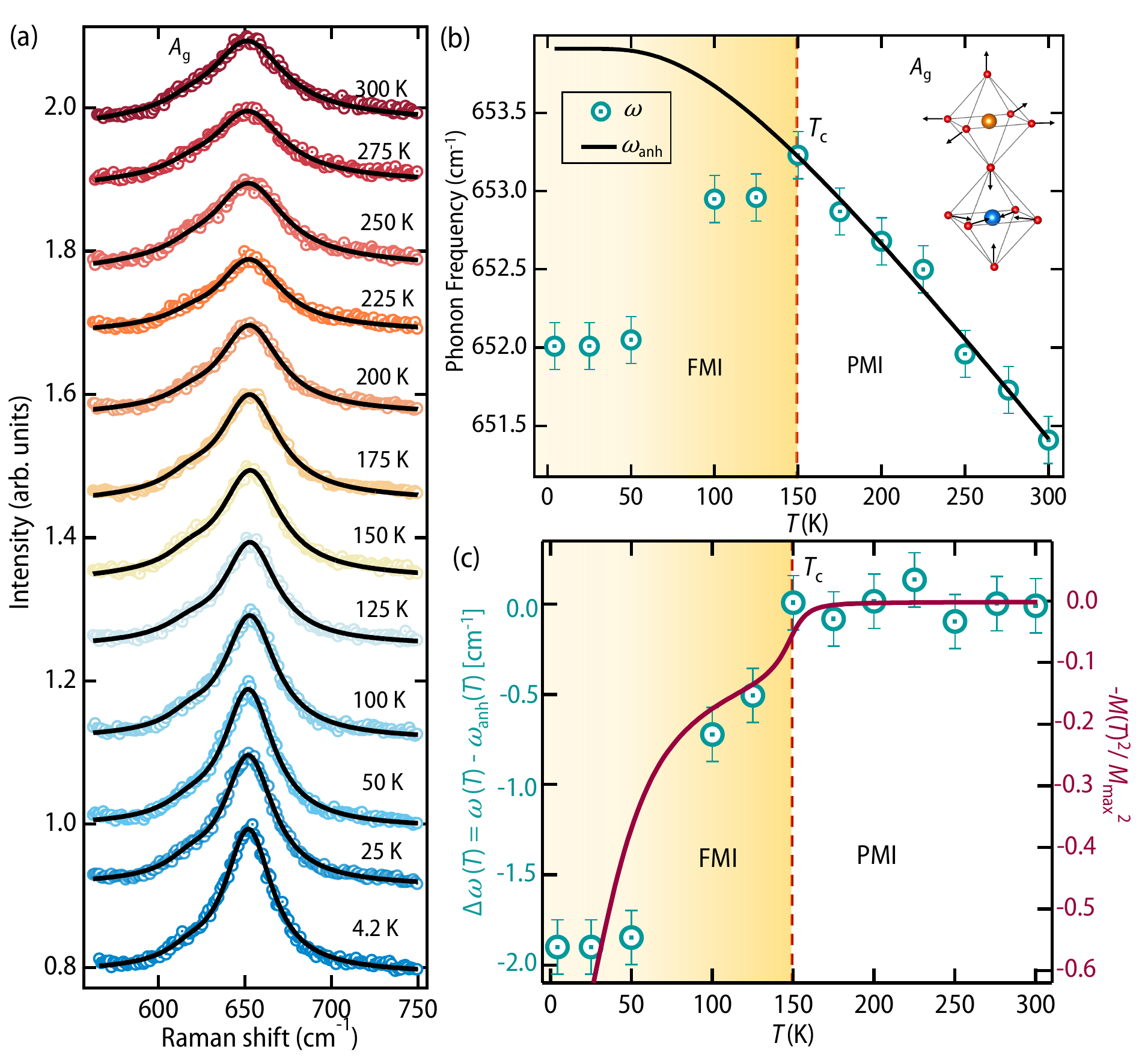}
	\caption{\label{Fig3} {\bf Temperature dependence of $A_\mathrm{g}$ raman mode:} (a)  Temperature-dependent Raman spectra of $RE^5$NMO measured from 4.2 K to 300 K, showing the evolution of the $A_\mathrm{g}$ phonon mode. 
    (b) Temperature dependence of the $A_\mathrm{g}$ mode peak position, fitted using an anharmonic phonon decay model (black curve). FMI and PMI represent ferromagnetic insulating and paramagnetic insulating states respectively.
    (c) Comparison of phonon frequency shift $\Delta \omega(T)$, obtained by subtracting the peak positions derived from Lorentzian fitting, from the values obtained from anharmonic model, with $M^2(T)/M_{\mathrm{max}}^2$.}

\end{figure}

Since the $B_g$ Raman mode $\sim$ 510 cm$^{-1}$ has strong contribution from the STO substrate with increase in temperature, we focus on the analysis of Raman mode around 653 cm$^{-1}$ [Fig.~\ref{Fig3}(a)],  which is assigned to the symmetric $A_\mathrm{g}$ stretching vibration of the $B$O$_6$ octahedra ($B$ = Ni, Mn) and is highly sensitive to changes in the lattice dynamics and magnetic order~\cite{Iliev:2007p104118, Iliev:2007p104118, Laverdiere:2006p214301, Macedo:2015p075201}.
  The spectra were fitted with two Lorentzian functions and the fitted curves are shown in Fig.~\ref{Fig3}(a): one to capture the $A_\mathrm{g}$ Raman mode, and another smaller peak at a lower frequency $\sim$ 618 cm$^{-1}$ to capture the monoclinic distortion~\cite{Macedo:2015p075201} and/or contribution from STO substrate since both overlap at similar phonon frequencies. The extracted $A_\mathrm{g}$ mode peak phonon frequencies are plotted as a function of temperature [Fig.~\ref{Fig3}(b)]. Further analysis of this mode revealed that the mode frequency above $T_\mathrm{c}$ fit well with the standard Balkinski model for anharmonic temperature dependence given by,
\begin{equation}
\omega(T) = \omega_0 - C \left(1 + \frac{2}{e^{\hbar\omega_0/2k_BT} - 1} \right)
\end{equation}
where $\omega$($T$) is the phonon frequency at temperature $T$, $\omega_0$ is the phonon frequency at $T = 0$ K and $C$ is the anharmonic constant~\cite{Iliev:2007p104118, Haro:1986p5358}. A stark deviation from the anharmonic model and softening of the $A_\mathrm{g}$ phonon mode is observed below 150 K that corresponds to the  $T_\mathrm{c}$ found from our magnetization measurements. Such softening has been shown to emanate from magnetic order induced phonon-renormalization, a clear indicator of strong spin-phonon coupling in the system in the magnetically ordered phase, as has been also observed in the parent $RE_2$NiMnO$_6$ compounds~\cite {Iliev:2007p151914,Macedo:2015p075201}.

Within the molecular field approximation, such phonon renormalization is proportional to the spin-spin correlation function $\langle \vec{S}_i \cdot \vec{S}_j \rangle$ for nearest-neighbour localized spins~\cite{Granado:1999p11879}. In case of ferromagnetic interaction under mean field approximation, $\langle \vec{S}_i \cdot \vec{S}_j \rangle$ is proportional to $M(T)^2$ giving \(\Delta \omega(T) = \omega(T) - \omega_{\mathrm{anh}}(T)\) $\propto$ \(-M^2(T)/M_{\text{max}}^2\) ~\cite{Laverdiere:2006p214301,Iliev:2007p104118,Macedo:2015p075201}. As shown in Fig.~\ref{Fig3}(c), the phonon frequency shift \(\Delta \omega(T) = \omega(T) - \omega_{\mathrm{anh}}(T)\) is plotted alongside the normalized square of magnetization term, \(-M^2(T)/M_{\text{max}}^2\) which is derived from the $M-T$ measurements shown in Fig.~\ref{Fig2}(c). The close correspondence between these two curves around ($T_\mathrm{c}$) confirms that the anomalous phonon softening originates from the spin-phonon coupling in the ferromagnetic phase and further validates a mean field description of the magnetism~\cite{Iliev:2007p104118,Macedo:2015p075201,Truong:2009p134424}. Interestingly,  we observe a deviation from this mean-field behavior at lower temperatures, similar to the magnetic anomaly observed in the ZFC measurements. 

\begin{figure}
	\vspace{-2pt}
	\hspace{-2pt}
	\includegraphics[width=0.8\textwidth] {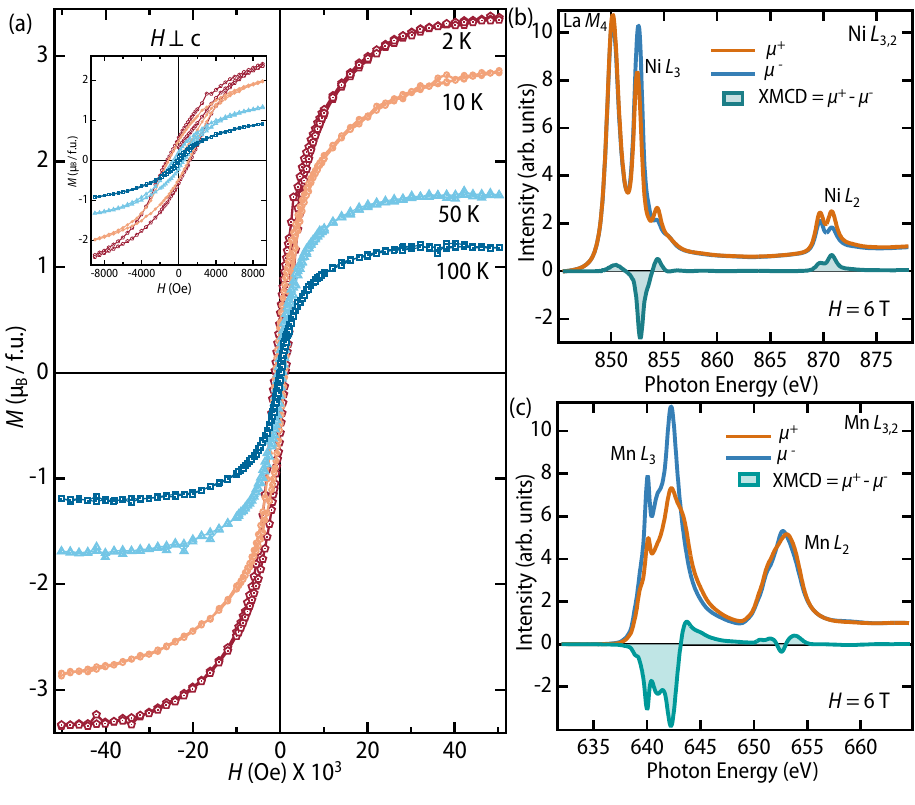}
	\caption{\label{Fig4}{\bf Ferromagnetic hysteresis and Ni/Mn coupling:} (a) $M$-$H$ curves  at 2 K, 10 K, 50 K, and 100 K for $H \perp c$. Inset: Corresponding zoomed $M$-$H$ curves highlighting the hysteresis. Right and left circularly polarized XAS spectra (denoted by $\mu^+$ and $\mu^-$ respectively) along with their difference XMCD signal at (b) Ni $L_\mathrm{3,2}$ and (c) Mn $L_\mathrm{3,2}$ edge exhibiting ferromagnetic coupling.
    }
\end{figure}

\begin{figure*}[]
	\vspace{-2pt}
	\hspace{-2pt}
	\includegraphics[width=1.0\textwidth] {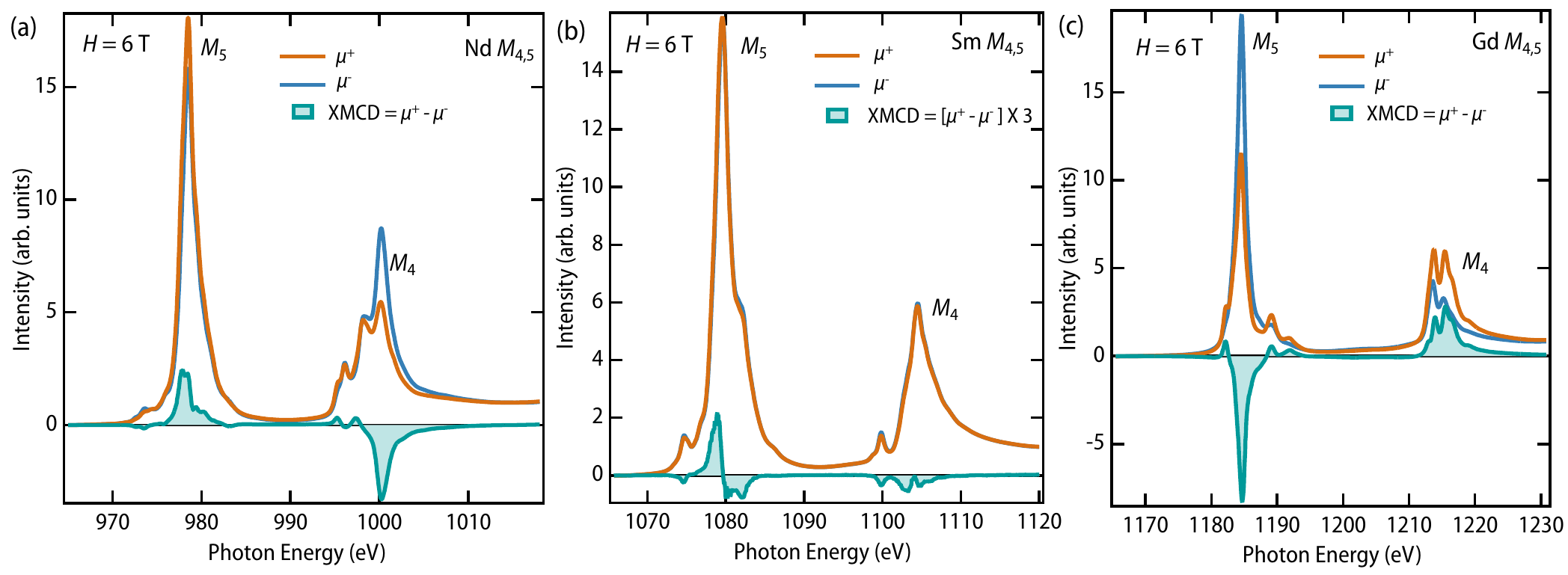}
	\caption{\label{Fig5} {\bf XMCD at $RE$ edges: } Right and left circularly polarized XAS spectra measured under 6 Tesla magnetic field for the $RE$ edges, along with their difference XMCD signal at (a) Nd $M_\mathrm{4,5}$, (b) Sm $M_\mathrm{4,5}$, and (c) Gd $M_\mathrm{4,5}$ edges.
    }
\end{figure*}

{\color{magenta}\bf Isothermal magnetization versus field and nature of magnetic couplings:}
We also performed isothermal $M–H$ measurements at selected temperatures below $T_\mathrm{c}$. Fig~\ref{Fig4}(a) displays $M-H$ loops at 2, 10, 50, and 100 K for \(H\perp c\). We observe a well-defined hysteresis loop consistent with the ferromagnetic behavior. However, we do not observe any significant perpendicular magnetic anisotropy in the presence case. Starting with the highest temperature 100 K [Fig.~\ref{Fig4}(a)], the $M-H$ curve exhibits soft ferromagnetism with a coercivity $\sim$ 200 Oe. The magnetization at higher fields shows a near saturation at about 1 $\mu_B$/f.u. (f.u. = formulae unit), and 1.6 $\mu_B$/f.u. for 100 K, and 50 K, respectively. Conventionally, on lowering the temperature, it is expected to be a clearer saturation of magnetization with an enhancement of the coercive field ($H_c$). While the $H_c$  increases to $\sim$ 1200 Oe for 2 K,  we find an increasing non-saturating trend of $M$ at higher $H$. To probe this peculiar behaviour, we perform element-specific X-ray magnetic circular dichroism (XMCD) measurements at both the TM and $RE$ cation edges, crucial for determination of the magnitude and alignment of their individual spin and orbital magnetic moments~\cite{Van:1986p6529,Vaz:2025p27}.

We first examine the coupling between the Ni and Mn sites from their respective XMCD signal obtained by the difference between the right and left circularly polarized absorption spectra ($\mu^+$ -$\mu^-$). All spectra were measured at the BOREAS beamline of the ALBA synchrotron, Spain, under a magnetic field of 6 T and temperature of 2 K, in grazing incidence geometry [See Methods for details]. The sign of the leading edge of the XMCD with the relatively larger spectral weight is dictated by the spin magnetic moment~\cite{Pal:2019p045122}. From the similar relative sign of the XMCD leading edge for both the Ni and Mn spectra [Fig.~\ref{Fig4}(b),(c)], we confirm the ferromagnetic coupling between the Ni and Mn cations akin to the parent compounds ~\cite{Pal:2018p165137, Pal:2019p045122,Spring:2023p104407}. Sum-rule analysis of the XMCD signal ~\cite{Carra:1993p694,Piamonteze:2009p184410} at the Mn $L_{3,2}$ edge yielded a spin moment $m_s \approx 2.11~\mu_B$ and an orbital moment $m_l \approx 0.08~\mu_B$, consistent with the expected quenching of orbital angular momentum in 3$d$ cations~\cite{Spring:2023p104407}. This $m_s$ is much smaller compared to the expected theoretical value of 3.87 $\mu_B$ for Mn$^{4+}$ [$S=3/2$]. The reduction is related to the Ni-Mn ASD, which  introduced  antiferromagnetic Mn$^{4+}$-O-Mn$^{4+}$ and  Ni$^{2+}$-O-Ni$^{2+}$ couplings~\cite{Pal:2018p165137,Majumder:2022p024408}. Due to the overlap with the La $M_4$ edge, our attempt to perform a similar analysis on the Ni $L_{3,2}$ XMCD signal resulted in large error bars, and  we could not get a reliable estimate of the Ni moments~\cite{Pal:2018p165137}.
 
 Fig.~\ref{Fig5}(a)-(c) show the XMCD signals for the $RE$ cations Nd, Sm and Gd at a magnetic field of 6 T at 2 K. The sign of the XMCD leading edges implies that the spin moments of Nd and Sm are aligned antiparallel, whereas those of Gd are aligned parallel to the Ni/Mn sublattice. 
However, it is unusual that, even in such high-field limit, Nd and Sm spins remain antiparallely aligned. We correlate this peculiarity to the fact that the effective magnetic moment for $RE$ atoms is the spin-orbit coupled moment. Since both Nd$^{3+}$ and Sm$^{3+}$ exhibit less than half-filled electronic configurations (Table~\ref{t1}), their total angular momentum is given by 
$J = |L - S|$. This yields a ground state in which the spin and orbital moments are mutually antiparallel, with the dominant orbital contribution leading to an overall parallel alignment with the Ni/Mn sublattice~\cite{Pal:2019p045122,Spring:2023p104407}. On the contrary, the orbital angular momentum $L$ of Gd$^{3+}$ is 0 (Table~\ref{t1}). Thus, the total moment of Gd$^{3+}$ has a spin-only contribution and exhibits a parallel alignment to the Ni/Mn sublattice as shown in Fig.~\ref{Fig5}(c). These consolidated results from XMCD further support the non-saturating nature of the $M-H$ plots in the high field limit, since with increasing $H$, the $RE$ cations begin to increasingly ferromagnetically align with the Ni/Mn sublattice. 

It is worth noting that while higher magnetic fields tend to drive the magnetic cations into parallel alignment, the behavior at lower fields reveals several intriguing features. 
In compounds of $RE_2$NiMnO$_6$ with $RE$ = Nd, Sm, when the applied field is weaker than a compensating threshold, it cannot overcome the internal field of the $RE$ sublattice, which aligns antiferromagnetically to the TM sublattice and produces a net decrease in magnetization with lowering of temperature ~\cite{Pal:2019p045122, Majumder:2022p094425}. Thus, investigating the low-field behavior of our system under such $RE$ site  compositional complexity will be particularly interesting.

\begin{figure} [ht!]
	\vspace{-2pt}
	\hspace{-2pt}
	\includegraphics[width=0.75\textwidth] {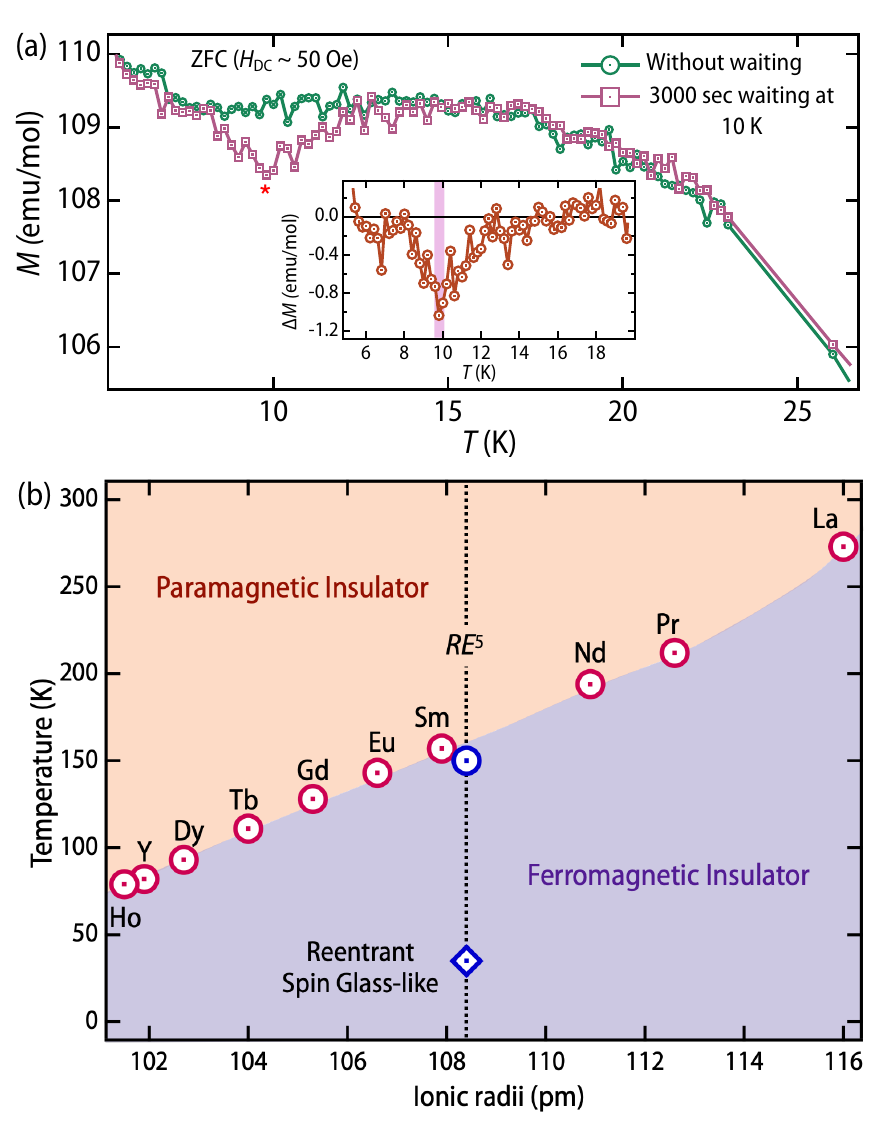}
	\caption{\label{Fig6}{\bf Memory experiment and magnetic phase diagram:} (a) ZFC magnetization recorded under $H_\mathrm{DC}$ = 50 Oe with and without halts at $T$ = 10 K. Inset: Corresponding difference curves $\Delta M$ versus $T$ (ZFC curve with halt minus curve without halt). (b) Magnetic phase diagram for $RE_2$NiMnO$_6$ compounds ($RE$ = La, Pr, Nd, Sm, Eu, Gd, Tb, Dy, Y and Ho) highlighting the paramagnetic to ferromagnetic transition along with the results from our work for single crystalline $RE^5$NMO. $T^*$ also has been additionally shown. 
    }
\end{figure}

{\color{magenta}\bf Reentrant spin glass-like behavior:} We next examine the low field, low temperature regime, which also indicated towards the possibility of a reentrant spin-glass state~\cite{Choudhury:2012p127201,Kumar:2014p011037} from our $M-T$ measurements [Fig.~\ref{Fig2}(c)-(e)]. To experimentally verify this, we performed memory experiments to capture any glassy behavior at lower temperatures under a small applied field.  When a spin glass system is quenched from a temperature above the glassy transition ($T_g$) down to a temperature lower than that, and then halted for a significant waiting time, aging effect is expected to be visible~\cite{Mathieu:2001p092401, Pramanik:2024p3382, Choudhury:2012p127201}. Following this protocol, the zero-field-cooled (ZFC) magnetization was first recorded during warming in a small magnetic field ($H_\mathrm{DC}$ $\sim$ 50 Oe), following a continuous cooling from a temperature well above $T^*$ down to the base temperature. A second ZFC measurement was then performed with an intermediate halt at $\sim$ 10 K, i.e. below $T^*$, where the sample was held for a halt time of 3000 sec before resuming the cooling [Fig.~\ref{Fig6}(a)]. In both protocols, the magnetic field remained zero throughout the cooling and waiting processes. A pronounced cusp appears near 10 K in the aged curve (which is the halting temperature marked by *) in comparison to the non-aged curve. The inset shows the difference curve given by, \(\Delta M = M_{\mathrm{3000 sec}} - M_{\mathrm{no wait}}\) showing a `memory dip' [Inset Fig.~\ref{Fig6}(a)]. This irreversibility behavior from aging experiments is a hallmark signature for spin glass states~\cite{Mathieu:2001p092401,Pramanik:2024p3382,Choudhury:2012p127201}. Since the system is magnetically ordered at higher temperature ($T_\mathrm{c}$), the data implies a reentrant spin-glass-like phase below $T^*$~\cite{Choudhury:2012p127201, Kumar:2014p011037}.

{\color{magenta}\bf Discussions:}
We explore the potential sources of magnetic frustration that lead to the reentrant spin-glass-like phase. The presence of ASD in $RE_2$NiMnO$_6$ introduces additional antiferromagnetic couplings, specifically through Ni$^{2+}$–O–Ni$^{2+}$ and Mn$^{4+}$–O–Mn$^{4+}$ superexchange interactions~\cite{Choudhury:2012p127201,Pal:2018p165137}. These competing interactions coexist with the dominant ferromagnetic Ni$^{2+}$–O–Mn$^{4+}$ superexchange. This competition between magnetic interactions is believed to be the origin of the reentrant spin-glass phase, which appears at approximately 100 K in La$_2$NiMnO$_6$ and Nd$_2$NiMnO$_6$ samples with significant ASD~\cite{Shi:2011p245405,Iliev:2009p023515,Choudhury:2012p127201,Singh:2016p74}. 
At this temperature scale ($\sim$ 100 K), the magnetic interaction between the Ni/Mn sublattice and the Nd ions is considered negligible, as evidenced by the consistent temperature scale of this phase in both compounds, given that La is nonmagnetic. Additionally, La$_2$NiMnO$_6$ and Nd$_2$NiMnO$_6$ samples, prepared with very little ASD, do not exhibit spin glass state~\cite{Pal:2019p045122,Yang:2012p064104,Sanchez:2011226001}. On the other hand, the glassy phase appearing around 20 K in Sm$_2$NiMnO$_6$ has been described by the magnetic coupling of Sm$^{3+}$ ion with Ni/Mn sublattice~\cite{Lekshmi:2013p6565}. Furthermore, ASD induced spin glass states are often accompanied by exchange bias effects due to spin-glass/ferromagnetic phase coexistence~\cite{Wang:2009p252502,Singh:2017p102402}.  Significantly, no exchange bias was observed in the present study. This finding indicates the absence of magnetic phase separation, thereby suggesting that the spin-glass signature originates from the entire sample volume.

Interestingly, our $M$–$T$ measurements reveal dynamical magnetic features below $\sim$ 35 K, comparable to the $RE$–sublattice interaction temperature with the Ni/Mn sublattice. This suggests a second scenario in which disorder at the $RE$ site adds further complexity to both the nature and strength of the magnetic interactions. Given that the memory experiments were conducted at a low field of 50 Oe, any field-induced parallel alignment of Nd and Sm is unlikely. Instead, under such conditions, their net moments align antiferromagnetically with the Ni/Mn sublattice, as revealed by earlier reports on neutron diffraction on parent compounds~\cite{Sanchez:2011226001,Majumder:2022p094425}. This coupling further manifests as a downturn in magnetization at low temperatures in Nd$_2$NiMnO$_6$ and Sm$_2$NiMnO$_6$ ~\cite{Pal:2019p045122,Majumder:2022p024408}. 
However, in our system, such a downturn trend is overcome by the strong paramagnetic response of the Gd, leading to an upturn in magnetization with lowering of temperature~\cite{Booth:2009p1559}.  Overall, the $RE$ sublattice via its coupling to the Ni/Mn network creates a complex landscape of competing magnetic interactions,  varying in both nature and strength.

While both of the above mechanisms may jointly contribute to the glassy behavior, the markedly lower transition temperature observed here differs from the values reported for purely ASD-driven spin-glass states in the parent compounds. Fig.~\ref{Fig6}(b) summarizes our observations, showing that the ferromagnetic transition in our system aligns with the general magnetic phase diagram of $RE_2$NiMnO$_6$ compounds. The average ionic size remains a reliable predictor of $T_\mathrm{c}$ even in this high variance compositionally complex settings, while a glassy phase appears at lower temperatures coinciding with the temperature scale of $RE$–sublattice interactions.

{\bf Conclusions and Outlook:} In this study, we demonstrated that long-range ferromagnetic order can robustly persist in a high variance compositionally complex DPO (La$_{0.4}$Nd$_{0.4}$Sm$_{0.4}$Gd$_{0.4}$Y$_{0.4}$)Ni-MnO$_{6}$. Our results demonstrate that the ferromagnetic transition temperature is primarily governed by the average ionic radius rather than the degree of variance, aligning with mean-field scenario. 
Below $T_\mathrm{c}$ , the phonon frequency does not follow the expected anharmonic trend, instead a mean-field-based phonon renormalization prevailed. However, a deviation from this mean-field behavior and the emergence of a reentrant spin glass-like anomaly were observed, which was further confirmed by magnetic memory experiments. Such frustration may arise from both Ni/Mn antisite disorder and $RE$-site disorder; however, the temperature of appearance of spin glass-like anomaly is closer to the $RE$–Ni/Mn interaction scale which points towards the $RE$-site exchange disorder as a possibly dominant driver in this case. 

Overall, this work opens a pathway for designer ferromagnetic high-entropy materials for the targeted magnetic transition temperature.
Considering the known appearance of a ferroelectric phase at the $T_\mathrm{c}$ in Y$_2$NiMnO$_6$~\cite{Su:2015p13260}, future research could focus on investigating the local structure of $RE^5$NiMnO$_6$ using advanced techniques like electron microscopy~\cite{Bhattacharya:2025p2418490}. This would provide valuable insights into the potential for magnetoelectric and multiferroic functionalities of this new class of materials.

\section*{Methods}

{\bf Sample preparation and characterization:} Polycrystalline target of $RE^5$NMO  was synthesized using solid state synthesis reaction by mixing stoichiometric quantities of La$_2$O$_3$, Nd$_2$O$_3$, Sm$_2$O$_3$, Gd$_2$O$_3$, Y$_2$O$_3$, NiO and MnO$_2$. The mixture was heated multiple times with intermediate grinding with a final heating at 1300 °C. The powder was pressed into a pellet and sintered to yield the target for PLD growth.

Single crystalline films of $RE^5$NMO (thicknesses 6 and 100 nm) were grown on TiO$_2$-terminated SrTiO$_3$ [001] substrates acquired from Shikosha, Japan, using a Neocera-based PLD system. A KrF excimer laser operating at $\lambda$ =
248 nm with a fluence of 2 J/cm$^2$ and repetition rate 2 Hz
was used and the growth was monitored via an \emph{in-situ} reflection high energy electron diffraction (RHEED) setup from Staib instruments. The films were grown at a deposition temperature of 750 $^{\circ}$C at a partial oxygen pressure of 150 mTorr. The films were annealed post-growth at 500 Torr for 30 minutes at the deposition temperature. 

Post-characterization of the films were performed with XRD and XRR measurements using a lab-based Rigaku X-ray diffractometer and the XRR fitting was carried out using GenX software~\cite{Bjorck:2007p1174}. Diffuse reflectance spectra of the film was recorded using a Perkin Elmer LAMBDA UV–Vis spectrophotometer to determine the optical band gap.

{\bf XAS and XMCD measurements:} XAS measurements at Ni $L_{3,2}$, Mn $L_{3,2}$ edges were carried out at beamline P04 PETRA III, DESY, Hamburg, Germany. The spectra were collected at room temperature in surface sensitive total electron yield (TEY) mode at grazing incidence geometry (angle of incidence $\sim$ 20$^\circ$). XMCD measurements with circularly polarized X-rays were performed at the Ni $L_{3,2}$, Mn $L_{3,2}$, Nd $M_{4,5}$, Sm $M_{4,5}$ and Gd $M_{4,5}$ edges at the BOREAS beamline, ALBA, Barcelona, Spain. The spectra were recorded at 2 K under the TEY mode. A magnetic field of 6 Tesla was applied parallel to the incident beam at grazing incidence. 

{\bf Magnetic measurements:} DC magnetization measurements were carried out using an MPMS XL superconducting quantum interference device (SQUID) magnetometer from Quantum Design Inc. 

{\bf Raman measurements:} Temperature-dependent Raman spectroscopy measurements were performed using Oxford-WITec Alpha 300R confocal photoluminescence Raman spectromicroscope in the 4 K to 300 K temperature range using liquid Helium. A laser wavelength of 532 nm and 1800 rules/mm grating was used to record the spectra in the wavenumber range of 100 to 850 cm$^{-1}$.

\section*{Acknowledgements}
 NB and SM thank Dr. Subhadeep Datta for insightful discussions on Raman spectroscopy data analysis. The authors acknowledge the use of central facilities of the Department of Physics, IISc, funded through the FIST program of the Department of Science and Technology (DST), Gov. of India.  SM acknowledges funding support from a
SERB Core Research grant (Grant No. CRG/2022/
001906) and a DST Nano Mission consortium project (No. DST/NM/TUE/QM-5/2019). RKD and RM thank the Olle Engkvist Stiftelse (Grant No. 224-0046) for financial
support. NB, SD acknowledge support from the Prime Minister’s Research Fellowship (PMRF), MoE, Government of India. NB also thanks Shakipriya Patra for assistance with solid-state synthesis.

\end{document}